# A space-resolved visible spectrometer system using compact endoscopic optics for full vertical profile measurement of impurity line emissions in superconducting EAST tokamak

A. Hu,[a] Y. Cheng,[a] L. Zhang,[a1] S. Morita,[ab] J. Ma,[ac] M. Kobayashi,[b] C. Zhou,[a] J. Chen,[a] Y. Cao,[ad] F. Zhang,[ac] W. Zhang,[ac] Z. Li,[ad] D. Mitnik,[e] S. Wang,[a] Y. Jie,[a] G. Zuo,[a] J. Qian,[a] H. Liu,[a] G. Xu,[a] J. Hu,[a] K. Lu,[a] Y. Song [a]

[a] *Institute of Plasma Physics Chinese Academy of Sciences, Hefei 230031, China*
[b] *National Institute for Fusion Science, Toki, 509-5292, Japan*
[c] *University of Science and Technology of China, Hefei 230026, China*
[d] *Anhui University, Hefei 230601, China*
[e] *Instituto de Astronomúy Física del Espacio (CONICET-Universidad de Buenos Aires), Buenos Aires 1428, Argentina*

[1]E-mail: zhangling@ipp.ac.cn

ABSTRACT: In Experimental Advanced Superconducting Tokamak (EAST tokamak) with tungsten divertors and molybdenum first wall, lithiumization and boronization have been frequently carried out to improve the plasma performance, in particular, in long pulse discharges. A study on impurity behaviors of lithium, boron and tungsten atoms/ions in the edge plasma is then crucially important. For the purpose, a space-resolved visible spectrometer system has been newly developed to observe full vertical profiles over a length of 1.7m of impurity line emissions in wavelength range of 320-800nm. For the full vertical profile measurement compact endoscopic optics is employed with an optical fiber bundle for the system, which can be inserted into a 1.5m long extension tube called 'long nose', because the distance between the diagnostic port and plasma center is considerably long (~2m). Therefore, a quartz glass window mounted from the vacuum vessel side is designed to withstand the reverse pressure. A mechanical shutter is also designed to open at a large angle of 235 degree so that the viewing angle of nearby ports is not blocked. Two sets of the fiber bundle, 60-channel linear array (core diameter: 115μm) and 11×10 channel planar array (core diameter: 50μm), with a length of 30m are attached to two sets of Czerny-Turner visible spectrometers for one-dimensional (1D) vertical profile measurement of core plasma and two-dimensional (2D) spectroscopy of divertor plasma, respectively. A complementary metal oxide semiconductor (CMOS) detector with 2048×2048 pixels (11μm/pixel) is used for the visible spectrometers. A preliminary result on the full vertical profile is obtained for BII line emission at 703.19nm in the 1D system.

Corresponding author: Ling ZHANG

Contents

## 1. Introduction

In the superconducting EAST tokamak with molybdenum first wall and tungsten divertors the long pulse discharge operation is the most important research subject. Recently, a thousand-second time scale (~1056s) long pulse discharge with Super I-mode has been achieved [1]. Lithiumization and boronization have been frequently carried out to sustain the long pulse discharge and to improve the plasma performance. A study on impurity behaviors of lithium, boron and tungsten atoms/ions in edge and divertor plasmas is then crucially important. Visible spectroscopy has been generally used as a tool for the study.

    The visible spectrometer system with optical fiber array has been installed on several tokamaks to observe radial profiles of hydrogen and impurity line emissions [2-7]. When the distance between the diagnostic port and the toroidal plasma is short, the fiber optics for the radial profile observation can be easily designed. In the superconducting tokamak like EAST, however, the distance is long (~2m) and the plasma cross section is vertically elongated. Therefore, it is difficult for EAST to measure the full-vertical profile of impurity line emissions in the usual way. Visible spectroscopy at superconducting LHD device with a long port−plasma dictance (~3m) is the only exception because the diagnostic port is very large and the vertical port size is a little larger than the vertical plasma size. In LHD spectroscopy, then, parallel fiber optics could be employed for the full-vertical profile measurment [8]. Here, it should be noted that the EAST tokamak has a double-null configuration and there are no diagnostic ports on the top and bottom of the vacuum vessel. Therefore, the horizontal profile measurement from the top or bottom port is imporssible, while the full-horizontal profile is measued in LHD using fun-array fiber optics [9]. To overcome these difficulties an alternative technique has been developed using compact endoscopic optics and a long-nose extention tube to observe the full radil profile. A preliminary result on the full vertical profile measurement is obtained for BII line emission at 703.19nm.

## 2. Visible Spectroscopy with Long-Nose Extension Tube

An overall schematic of the visible spectroscopic system is shown in figure 1. Two sets of the fiber bundle are installed in the vicinity of the plasma to measure 1D full-vertical profile of the core plasma and 2D distribution of the upper tungsten divertor plasma. The field of view (FOV) of 1D and 2D systems are indicated in figure 1. The two fiber bundles with the lenz optics are combined as fiber bundle module and inserted into a long-nose extension tube. Visible light collected by lens optics is transferred through the 30m long fiber bundle to the visible spectrometer located in the basement room. The 30m long UV-grade fiber allows the transmission rate of 33% at 300nm, 72% at 400nm and 87% at 500nm. These details are described below

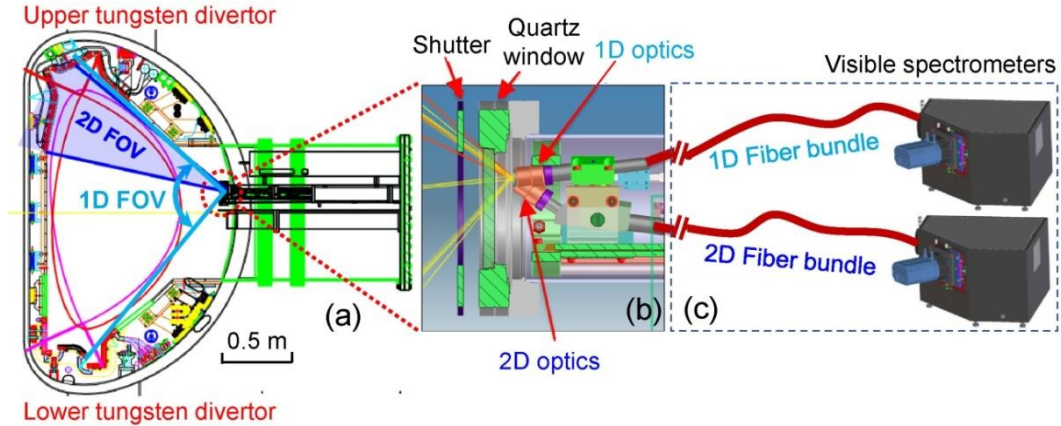

Figure 1. Diagrammatic sketch of visible spectrometer system; (a) EAST poloidal cross section at F port with field of view (FOV) for one-dimensional (1D) and two-dimensional (2D) visible spectroscopy, (b) detail of endoscopic optics in the long-nose extension tube and (c) visible spectrometers with 30m long fiber bundle for core (1D) and edge (2D) plasma diagnostics.

## 2.1 Shutter and Quartz Glass Window

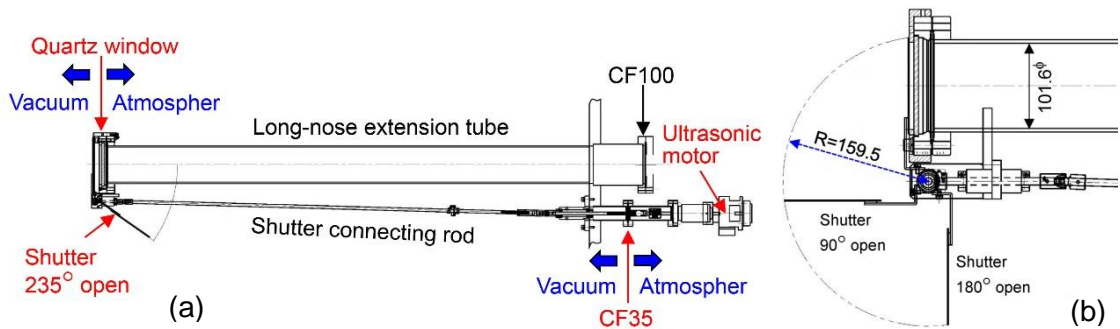

Figure 2. Schematic drawings of (a) long-nose extension tube with quartz window and mechanical shutter and (b) detail of the shutter system.

The long-nose extension tube with a length of 1.5m is shown in figure 2 with a mechanical shutter for protection of the glass window. Using the extension tube in atmospheric pressure the fiber bundle can be installed very close to the plasma. The shutter plate made of stainless steel can be mechanically rotated by an ultrasonic motor through a shutter connecting rod, universal joint and gears. Since the motor is located close to magnetic coils, the ultrasonic motor is selected in the present system. The shutter connecting rod attached on a CF35 conflat flange is fixed on the vacuum side to the extension tube to prevent it from vibrating. The shutter plate can be rotated at a large angle so that it does not interfere with the view of neighboring diagnostics. In daily experiments, the shutter plate is fixed at 235 degrees open, as shown in figure 1(a).

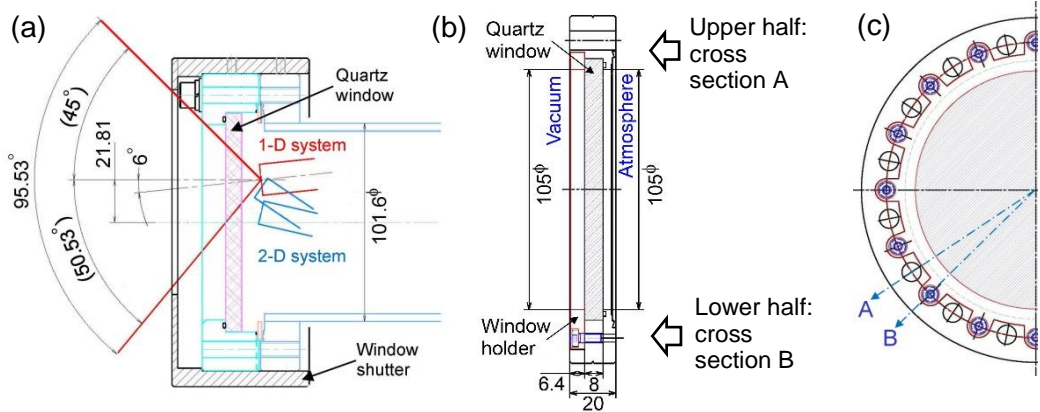

Figure 3. (a) Arrangement of two fiber optics and glass window, (b) cross-sectional drawing of quartz glass window structure for vacuum-atmosphere reverse use, (c) structure of CF100 conflat flange. Upper and lower half drawings in (b) indicate cross sections along A and B lines in (c), respectively.

Structure of the glass window is explained in figure 3. The two fiber bundle positions and shutter opening area shown in figure 3(a) are carefully desigend to enable the full vertical profile measurement by the 1D system and to ensure a sufficient area for divertor spectroscopy by the 2D system. The glass window made of quartz is used with reverse pressure as shown in figure 3(b). For the purpose the glass window is tightly fixed to the back of the CF100 conflat flange by the holder via 16 screws. The additional 16 screws are seen in figure 3(c) at a line of B. A rubber ring with square cross-sectional shape is used for complete vacuum sealing. Then, the leak rate of the glass window is kept below $10^{-11}$ torr·l/s. The EAST experiment was continued for two years after installation of the glass window, but no vacuum leakage from the glass window occured.

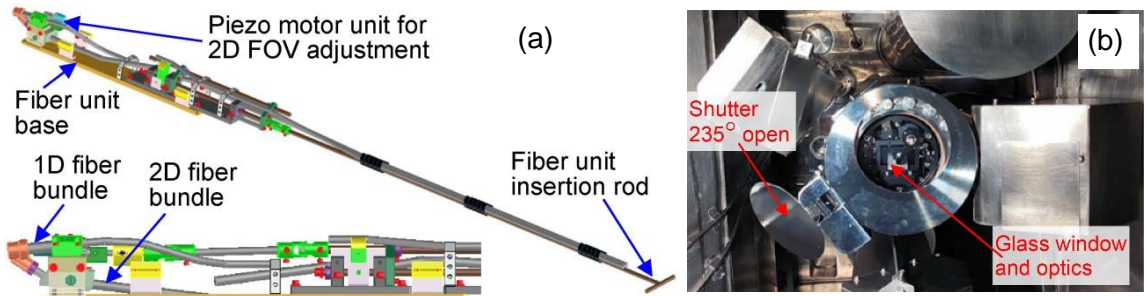

Figure 4. (a) Fiber bundle module for insertion into the long-nose extension tube and (b) photo of glass window with mechanical shutter taken from inside EAST vauum vessel.

## 2.2 Fiber Bundle Module

The fiber bundle module is explained in figure 4(a). The two fiber bundles of 1D and 2D systems are compactly mounted on a fiber unit base. The angle of the lenz optics at the 2D fiber bundle head can be vertically tilted by a Piezo motor unit for the FOV adjustment as shown in figure 3(a), while the 1D fiber bundle head is fixed. The angle adjustment can be made both by computer control and manual. The module can be easily removed by pulling a fiber unit insertion rod by hand. A photograph taken from inside the vacuum vessel is shown in figure 4(b) with glass window, fiber optics and 235° opened shutter.

The 1D fiber bundle consists of 60 linearly arranged optical fibers with a quartz core of 115μm diameter and a clad of 125μm diameter. The 2D fiber bundle totally consists of 110 optical fibers with a quartz core of 50μm diameter and a clad of 62.5μm diameter. The fibers are arranged in 10×11 rectangle at the lenz optics to observe the 2D image of the divertor plasma and liniarly arranged at the entrance slit of the visible spectrometer for visible spectroscopy. Based on the module, the full-vertical profile of impurity line emissions can be observed over a length of 1.7m in the plasma center uisng the 1D system and a divertor area of 310×340mm$^2$ can be covered for divertor spectroscopy using the 2D system.

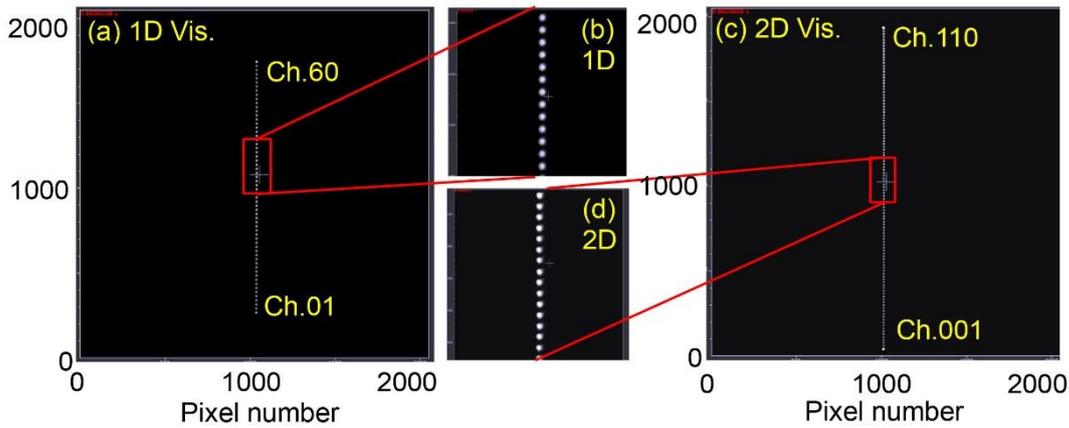

Figure 5. Example of focal image adjustment results obtained with CMOS detector and neon light source at (a) full-size (2024×2024 pixels) and (b) enlarged size (pixel no. 890-1220) in 1D spectrometer system with 60 fiber channels and (c) full-size (2024×2024 pixels) and (d) enlarged size (pixel no. 850-1150) in 2D spectrometer system with 110 fiber channels

**2.3 Visible Spectrometer and Detector**

A Czerny-Turner type visible spectrometer with focal length of 280mm (MK300, Bunkoukeiki co. ltd) is used for both 1D and 2D spectrometers. The astigmatism is corrected by installing a toroidal mirror in addition to two spherical mirrors and a flat mirror. Three gratings (300, 1200 and 2400 grooves/mm) mounted in the spectrometer can be externally selected according to the experimental scenario. A complementary metal oxide semiconductor (CMOS: Andor Marana-4BV11) detector is attached to the spectrometer. The pixel size is 11×11μm and the total detector size is 22.528×22.528mm (2048×2048 pixels). Visible spectra in wavelength range of 200-1000nm can be measured in principle, although the short wavelength side is limited by the transmission rate of the optical fiber. The spectral resolution is roughly 0.2nm at the 1200 grooves/mm grating.

The angle and position of the CMOS detector and the optical fiber array mounted at the entrance slit were adjusted using a neon light source to achieve an optimal focused image. An example on the focal image adjustment result is shown in figure 5 for the 1D and 2D systems. The fiber images of 1D and 2D systems are both focused almost in a straight line as shown in figures 5(a) and (c), respectively. Magnified views of the 1D and 2D focal images are also shown in figures 5(b) and (d), respectively. It can be seen that each focal image almost has a circular shape reflecting the fiber geometry. These results indicate that the aberration and astigmatism are well corrected in the present visible spectrometer.

## 3. Preliminary Results

The first resuls were obtained from the 1D spectrometer system. Visible spectrum observed with 300grooves/mm grating is plotted in figure 6(a) in the wavelength of 375-725nm. The throughput of the spectrometer is absolutely calibrated using a standard integral sphere. Many visible spectral lines emitted from imurity and deuterium atoms/ions are identified in the spectrum. Most of the spectral lines originate in the light impurities. Spectral lines from only a metallic impurity are found at 388.85 and 541.10nm as FeI.

Time evolution of the full-vertical profile of BII line at 703.19nm is shown in figure 6(b). The profile observed from lower-divertor single-null discharge ($n_e=2.1\times10^{13}$cm$^{-3}$ and $T_e=0.9$keV) is obtained every 100ms. At the beginning of the discharge, t=0.2-0.9s, the BII emission appears only in the plasma central region, reflecting a circular plasma with small size. After the electron cyclotron heating starts at t=1.0s, the plasma quickly expands toward the divertor region. The BII emission reappears in the lower divertor region near Z=-40cm. The two separated BII emissions are originated from the distribution of magnetic field lines in the divertor region as shown in figure 1(a). Time-varying distribution of the BII emission seems to be due to positional and performance changes in the edge plasma.

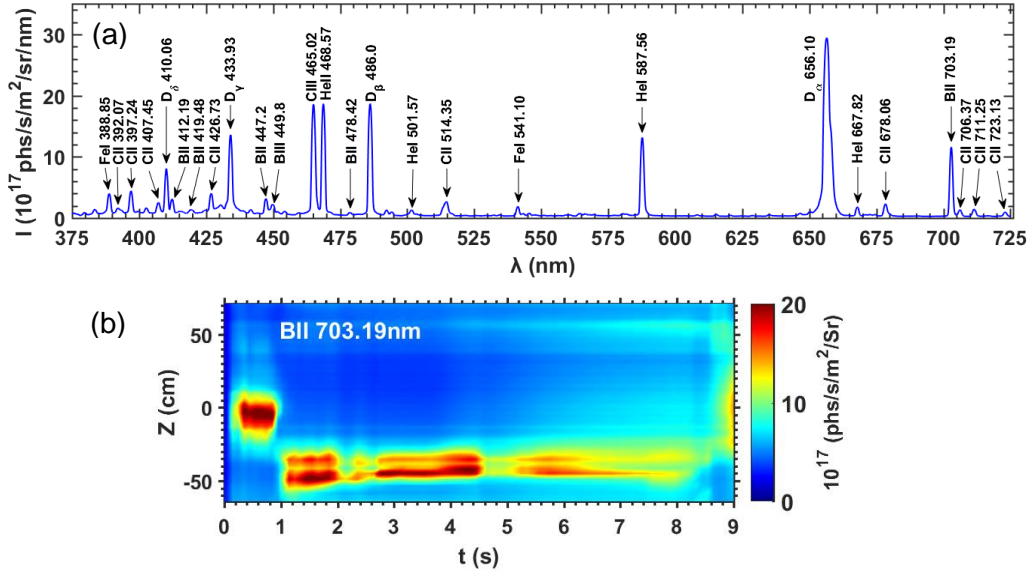

Figure 6. (a) Visible spectrum in wavelength range of 375-725nm at 300g/mm grating and (b) time evlution of BII vertical profile observed from discharge after boronization.

## 4. Summary

A space-resolved visible spectrometer system was newly constructed to observe full vertical profiles over a vertical length of 1.7m of impurity line emissions in the wavelength range of 320-800nm. For the full vertical profile measurement compact endoscopic optics inserted into a 1.5m long extension tube was developed with a quartz glass window mounted from the vacuum vessel side and a mechanical shutter which can be opened at a large angle of 235 degree. Two sets of the 30m long fiber bundle are attached to two sets of Czerny-Turner visible spectrometers for 1D and 2D spectroscopic diagnostics. The CMOS detector with 2048×2048 pixels is used for recording the visible spectra. A good focal image of the fiber optics was obtained after detailed adjustment of the spectrometer. A full vertical profile of BII line at 703.19nm was successfully observed from the 1D spectrometer system.

In the near future attempts are made to observe visible spectra from low-ionized and highly charged tungsten ions and radial profiles of effective ion charge, $Z_{eff}$, using the 1D spectrometer system.

## Acknowledgments

This work was supported by the National Magnetic Confinement Fusion Energy R&D Program of China (Grant Nos. 2022YFE03180400, 2019YFE030403, 2022YFE03020004, 2024YEF03000200), National Natural Science Foundation of China (Grant No. 12322512) and Chinese Academy of Sciences President's International Fellowship Initiative (PIFI) (Grant Nos. 2025PVA0060). We thank all members at EAST (https://cstr.cn/31130.02.EAST), for providing technical supports and assistances in the data collection and analysis.